   \documentclass[12pt]{article}
   \usepackage{amssymb}
   \makeatletter

   \@addtoreset{equation}{section}

   \makeatother

   \newtheorem{lemma}[equation]{Lemma}

   \renewcommand\thefigure{\thesection.\@arabic\c@figure}

   \renewcommand\thetable{\thesection.\@arabic\c@table}

   \def\reff#1{(\ref{#1})}
   \def\sobre#1#2{\lower 1ex \hbox{ $#1 \atop #2 $ } }

\begin{document}

   \def\E{{\mathbb E}}
   \def\P{{\mathbb P}}
   \def\R{{\mathbb R}}
   \def\Z{{\mathbb Z}}
   \def\V{{\mathbb V}}
   \def\N{{\mathbb N}}
   \def\B{{\mathbb B}}
   \def\bN{{\bf N}}
   \def\cX{{\cal X}}
   \def\cT{{\cal T}}
   \def\bC{{\bf C}}
   \def\bD{{\bf D}}
   \def\bG{{\bf G}}
   \def\bU{{\bf U}}
   \def\bK{{\bf K}}
   \def\bH{{\bf H}}
   \def\bS{{\bf S}}
   \def\cJ{{\cal J}}
   \def\bn{{\bf n}}
   \def\bd{{\bf d}}
   \def\bb{{\bf b}}
   \def\ba{{\bf a}}
   \def\ee{{\rm e}}
   \def\bg{{\bf g}}
   \def\mm{{m}}
   \def\ti{{\rm TI}}
   \def\sqr{\vcenter{
            \hrule height.1mm
            \hbox{\vrule width.1mm height2.2mm\kern2.18mm\vrule width.1mm}
            \hrule height.1mm}}                  % This is a slimmer sqr.
   \def\square{\ifmmode\sqr\else{$\sqr$}\fi}
   \def\one{{\bf 1}\hskip-.5mm}
   \def\liml{\lim_{L\to\infty}}
   \def\given{\ \vert \ }
   \def\Given{\ \Big\vert \ }
   \def\ze{{\zeta}}
   \def\be{{\beta}}
   \def\de{{\delta}}
   \def\la{{\lambda}}
   \def\ga{{\gamma}}
   \def\th{{\theta}}
   \def\vep{{\varepsilon}}
   \def\proof{\noindent{\bf Proof. }}
   \def\bA{{\bf A}}
   \def\bB{{\bf B}}
   \def\bD{{\bf D}}
   \def\bH{{\bf H}}
   \def\bh{{\bf h}}
   \def\bX{{\bf X}}
   \def\bY{{\bf Y}}
   \def\bZ{{\bf Z}}
   \def\bK{{\bf K}}
   \def\bF{{\bf F}}
   \def\bx{{\bf x}}
   \def\by{{\bf y}}
   \def\cw{{\cal W}}
   \def\zero{{\rm 0}}

   % New macros

   \def\VV{{\underline V}}
   \def\UU{{\underline U}}
   \def\XX{{\underline X}}
   \def\FF{{\cal F}}
   \def\GG{{\cal G}}
   \def\tends#1{\mathop{\longrightarrow}\limits_{#1}}% Arrow with limits
                                                     % below

   \title{On mixing times for stratified walks on the $d$-cube}
   \author{Nancy Lopes Garcia\\ Universidade Estadual de
     Campinas\\ Campinas, Brasil\\and\\Jos\'e Luis Palacios
   \\Universidad Sim\'on Bol\'{\i}var\\
   Caracas, Venezuela.}

   \maketitle

   \abstract{
   Using the electric and coupling approaches, we derive a series of
   results concerning the mixing times for the
   stratified random walk on the $d$-cube, inspired in the results of
   Chung and Graham (1997) Stratified random walks on the
   $n$-cube. {\it Random Structures and  Algorithms}, {\bf
   11},199-222.  
   }

   \vskip .2 in
   \noindent {\it Key Words:  effective resistance, coupling, birth and
     death chains}
   \vskip .2 in
   \noindent {\bf 1991 Mathematics Subject Classification.}  Primary:
   60J15; secondary: 60C05.
   \vfill
   \eject
   \section{ Introduction.}

   The stratified random walk (SRW) on the $d$-cube $Q_d$ is the Markov chain whose
   state space is the set of vertices of the $d$-cube and whose
   transition probabilities are defined thus:

   Given a set of non-zero probabilities $p=(p_0, p_1, \ldots, p_{d-1})$,
   from any vertex with $k$ 1's, the process moves either to any neighboring
   vertex with $k+1$ 1's with probability ${p_k\over d}$; or to any
   neighboring vertex with $k-1$ 1's with probability ${{p_{k-1}}\over
     d}$; or to itself with the remaining probability. The simple random
   walk on the $d$-cube corresponds to the choice $p_k=1$ for all $k$.

   Vaguely speaking, the mixing time of
   a Markov chain is the time it takes the chain to have its distribution
   close to the stationary distribution under some measure of closeness.
   Chung and Graham studied the SRW on the $d$-cube in  \cite{fc:rg:97},
   mainly with algebraic methods, and found bounds for the mixing
   times under total variation and relative pointwise distances.   Here we
   use non-algebraic methods,
   the electric and coupling approaches,  in order to study the same SRW
   and get exact results for
   maximal commute times and bounds for cover times and mixing times
   under total variation distance. We take advantage of the fact that
    there seems to be some
   inequality or another linking hitting times, commute times, cover
   times and any
   definition of mixing time with any
   other under any measure of closeness (see Aldous and
   Fill \cite{da:jf:99} and Lov\'asz and Winkler \cite{ll:pw:98}).

   \section{The electric approach}

   On a connected undirected graph $G=(V,E)$ such that the edge between
   vertices $i$ and $j$ is given a
   resistance $r_{ij}$ (or equivalently, a conductance $C_{ij}=1/r_{ij}$),
   we can define
    the random walk on $G$ as the Markov chain $\bX=\{\bX(n)\}_{n\ge 0}$ that from
   its current vertex $v$ jumps
   to the neighboring vertex $w$ with probability  $p_{vw}=C_{vw}/C(v)$,
   where $C(v)=\sum_{w: w\sim v} C_{vw}$, and $w\sim v$ means that $w$ is a
   neighbor of $v$.  There may be a
   conductance $C_{zz}$ from a vertex $z$
   to itself, giving rise to a transition probability from $z$ to
   itself.
     Some notation: $\E_aT_b$ and $\E_aC$ denote the expected value,
   starting from the
   vertex $a$, of respectively, the hitting time $T_b$ of the vertex $b$
   and the cover time $C$, i. e., the number of jumps needed to visit all
   the states in $V$; $R_{ab}$ is the effective resistance, as computed
   by means of Ohm's law, between vertices $a$ and $b$.

   A Markov chain is reversible if $\pi_i\P(i,j)=\pi_j\P(j,i)$ for all
   $i,j$, where $\{\pi_{.}\}$ is the stationary distribution and
   $\P(\cdot,\cdot)$ are the transition probabilities.  Such a reversible Markov
   chain can be described as a random walk on a graph if we define
   conductances thus:
   \begin{equation}
   \label{conduc}
   C_{ij}=\pi_i\P(i,j).
   \end{equation}

   We will be interested in finding a closed form expression for the
   commute time $\E_0T_d+\E_dT_0$ between the origin, denoted by 0,  and
   its opposite vertex, denoted by $d$.

   Notice first that
   the transition matrix for $\bX=\{\bX(n), n\ge 0\}$, the SRW on the $d$-cube, is doubly stochastic
   and therefore its stationary distribution is uniform.  If we now
   collapse all vertices in the cube with the same number of 1's into a
   single vertex, and we look at the SRW on this collapsed graph, we
   obtain a new {\it reversible} Markov chain $\bS=\{\bS(n), n\ge 0\}$, a birth-and-death chain in
   fact, on the state space $\{0, 1, \ldots d\}$, with transition
   probabilities

   %$$P(k, k+1)={{d-k}\over d}p_k,$$
   %$$P(k, k-1)={{k}\over d}p_{k-1},$$
   %$$P(k, k)=1-P(k,k+1)-P(k,k-1).$$
   \begin{eqnarray}
   \P(k, k+1)&=&{{d-k}\over d}p_k, \label{p:1} \\
   \P(k, k-1)&=&{{k}\over d}p_{k-1}, \label{p:2} \\
   \P(k, k)&=&1-\P(k,k+1)-\P(k,k-1). \label{p:3}
   \end{eqnarray}

   It is plain to see that the stationary distribution of this new
   chain is the Binomial with parameters $d$ and ${1 \over 2}$.  It is
   also clear that the commute time between vertices 0 and $d$ is
   the same for
   both $\bX$ and $\bS$.
   For the latter  we use the electric machinery described above,
    namely, we think of a linear electric circuit from $0$ to $d$ with
   conductances given by (\ref{conduc})
   for $0\le i \le d$, $j=i-1, i, i+1$, and where $\displaystyle \pi_i={d \choose
     i}{1\over 2^d}$.

   It is well known (at least since Chandra et al. proved it in  \cite{ac:etal:89}) that
   \begin{equation}
   \label{uno}
   \E_aT_b+\E_aT_b=R_{ab}\sum_zC(z),
   \end{equation}
   where  $R_{ab}$ is the effective resistance between vertices $a$ and $b$.

   If this formula is applied to a reversible chain whose conductances
   are given as in (\ref{conduc}), then it is clear that
   $$C(z)=\pi_z$$
   and therefore the summation in (\ref{uno}) equals 1. We get then this
   compact formula for the commute time:
   \begin{equation}
   \label{simple}
   \E_aT_b+\E_aT_b=R_{ab},
   \end{equation}
   where the effective resistance is computed with the individual
     resistors having resistances
   $$r_{ij}={1\over {C_{ij}}}={1\over {\pi_i\P(i,j)}}.$$

   In our particular case of the collapsed chain, because it is a linear
   circuit, the effective
   resistance $R_{0d}$ equals the sum of all the individual resistances
   $r_{i,i+1}$, so that (\ref{simple}) yields

   \begin{equation}
   \label{otra}
   \E_0T_d+\E_dT_0= R_{0d}=2^d\sum_{k=0}^{d-1}{1\over {p_k{{d-1}\choose
         k}}}.
   \end{equation}

   Because of the particular nature of the chain under consideration, it
   is clear that $\E_0T_d+\E_dT_0$ equals the maximal commute time
   ($\tau^*$ in the terminology of Aldous \cite{da:jf:99}) between any two vertices.

   (i) For simple random walk, formula (\ref{otra}) is simplified by
   taking all $p_k=1$.  This particular formula was obtained in
   \cite{ll:pw:98}{ with a more direct argument, and it was argued there that
   $$\sum_{k=0}^{d-1}{1\over {{d-1}\choose k}}=2+o(1).$$

   An application of Matthews' result (see \cite{pm:88}), linking maximum and
   minimum expected hitting times with expected cover times, yields immediately
   that the expected cover time is
   $\E_vC=\Theta(|V|\log |V|),$
   which is the asymptotic value of the lower bound for cover times of walks on a
 graph $G=(V,E)$ (see \cite{uf:95}).  Thus we could say this SRW is a ``rapidly
covered'' walk.

   (ii) The so-called Aldous cube (see  \cite{fc:rg:97}) corresponds to the choice
   $p_k={k\over {d-1}}$.  This walk takes place in the ``punctured cube''
   that excludes the origin.  Formula (\ref{otra}) thus,  must exclude
   $k=0$ in this case, for which we still get a closed-form expression for the
   commute time between vertex $d$, all of whose coordinates are 1,  and
   vertex $s$ which consists of the collapse of all vertices with a single 1:
   \begin{equation}
   \label{dos}
   \E_sT_d+\E_dT_s=2^d\sum_{k=1}^{d-1}{1\over {{d-2}\choose {k-1}}}.
   \end{equation}
   The same argument used in (i) tells us that the summation in
   (\ref{dos}) equals $2+o(1)$ and, once again, Matthews' result tells us
   that the walk on the Aldous cube has a cover time of order $|V|\log
   |V|$.

   (iii) The choice $p_k={1\over {{d-1}\choose k}}$ would be in the
   terminology of  \cite{fc:rg:97} a ``slow walk'': the commute time is seen to be
   exactly equal to $|V|\log_2 |V|$ and thus the expected cover time is
   $O(|V|\log^2|V|)$.  In general, the SRW will be rapidly covered if and
   only if
   $$\sum_{k=0}^{d-1}{1\over {p_k{{d-1}\choose k}}}=c+o(1),$$
   for some constant $c$.

   Remark.  A formula as compact as (\ref{otra}) could be easily obtained
   through the commute time formula (\ref{simple}).  It does not seem
   that it could be obtained that easily, by just adding the
   individual hitting times $\E_iT_{i+1}$. (A procedure that is done, for 
instance, in
    \cite{fc:rg:97}, \cite{jp:94}, \cite{jp:pt:96},  and in the next section).
   \vskip .1in

   \section{The coupling approach}

   In order to asess the rate of convergence of the SRW on the cube $Q_d$ to the
   uniform stationary distribution $\pi$, we will bound the mixing time $\tau$
   defined as
   $$\tau=\min\{t: d(t')\le \frac{1}{2e}, \mbox{ for all } t'>t\},$$
   where
   $$d(t)=\max_{\bx \in Q_d}\|P_{\bx}(\bX(t)=\cdot)-\pi(\cdot)\|,$$
   and $\|\theta_1-\theta_2\|$ is the variation distance between probability
   distributions $\theta_1$ and $\theta_2$, one of whose alternative definitions
   is (see Aldous and Fill \cite{da:jf:99}), chapter 2):
   $$\|\theta_1-\theta_2\|=\min \P(V_1\neq V_2),$$
   where the minimum is taken over random pairs $(V_1,V_2)$ such that $V_m$ has
   distribution $\theta_m, m=1,2.$

   The bound for the mixing time is achieved using a coupling argument that goes as follows: let
   $\{\bX(t), t \ge 0\}$ and $\{\bY(t), t \ge 0\}$ be two versions of the SRW on
   $Q_d$ such that $\bX(0)=\bx$ and $\bY(0)\sim \pi$. Then
   \begin{equation}
   \label{coupling1}
   \|\P_{\bx}(\bX(t) = \cdot) - \pi(\cdot)\| \le \P(\bX(t) \neq
   \bY(t)).
   \end{equation}

   A coupling between the processes $\bX$ and $\bY$ is a bivariate process such
   that its marginals have the distributions of the original processes and such that once the bivariate
    process enters the diagonal, it stays there forever. If we denote by
   \[ T_{\bx} = \inf\{t; \bX(t) = \bY(t)\}\]
   the coupling time, i. e., the hitting time of the diagonal, then
   (\ref{coupling1}) translates as
   \begin{equation}
   \label{classic}
    \|\P_{\bx}(\bX(t) = \cdot) - \pi(\cdot)\| \le \P(T_{\bx} > t),
   \end{equation}
   and therefore,
   \begin{equation}
   \label{coupling2}
   d(t) \le \max_{\bx \in Q_d} \P(T_{\bx} > t).
   \end{equation}

   If we can find a coupling such that $\E T_{\bx} =O(f(d))$,
   for all $x \in Q_d$ and for a certain function $f$ of the dimension $d$,
   then we will also have $\tau= O(f(d))$.  Indeed, if we
   take $t=2e f(d)$, then (\ref{coupling2}) and Markov's inequality
   imply that $d(t)\le 1/2e$ and the definition of $\tau$ implies
   $\tau=O(f(d))$.

   We will split $T_{\bx}$ as $T_{\bx}=T_{\bx}^1 + T_{\bx}^2$, where $T_{\bx}^1$
    is a coupling
   time for the birth-and-death process $\bS$, and $T_{\bx}^2$ is another
    coupling time for the whole
   process {\bX}, once the bivariate $\bS$ process enters the diagonal, and we
   will bound the values of $\E T_{\bx}^1$ and $ \E T_{\bx}^2$.

   More
   formally, for any $\bx, \by \in Q_d$ define
   \begin{eqnarray}
   s(\bx) = \sum_{i=1}^d x_i \label{n:2.1} \\
   d(\bx,\by) = \sum_{i=1}^d |x_i - y_i|. \label{n:2.2}
   \end{eqnarray}

   Define also for the birth-and-death process  $\bS(t)=s(\bX(t))$ its own mixing
   time:
   \[ \tau^{(S)} = \inf \{ t; d_{S}(t) \le \frac{1}{2e}\},\]
   where $d_S(t) = \max_i \|\P_i(\bS(t) = \cdot) - \pi_S(\cdot)
   \|$, and $\pi_S$ is the stationary distribution of $\bS$.
   Notice that $s(\bY(0)) \sim \pi_{S}$ since $\bY(0) \sim \pi$.

   Now we will prove that
   $ \tau^{(S)}=O(f_{S}(d)),$ for a certain function $f_S$ of the expected
   hitting times of the ``central states'',  and that this bound implies an
   analogous bound for $\E T_{\bx}^1$.
   Indeed, as shown by Aldous \cite{da:82}, we can bound $ \tau^{(S)}$ by a
   more convenient stopping time
   \begin{equation}
   \label{n:2.48}
    \tau^{(S)}  \le  K_2 \tau_1^{(3)}
   \end{equation}
   where $\tau_1^{(3)} = \min_{\mu} \max_i \min_{U_i} \E_i U_i$ and the
   innermost minimum is taken over stopping times $U_i$ such that
   $\P_i(S(U_i) \in \cdot) = \mu(\cdot)$. In particular,
   \begin{eqnarray}
   \tau_1^{(3)}  & \le & \min_b \max_i \min_{U^b_i} \E_i \, U^b_i
             \label{pa:1} \\
             & \le & \min_b \max_i \E_i T_b \label{n:2.49} \\
             & = &  \max(\E_0 T_{d/2}, \E_d T_{d/2} ) \label{n:2.49a}
   \end{eqnarray}
   where the innermost minimum in \reff{pa:1} is taken over stopping
   times $U^b_i$ 
   such that $\P_i(S(U_i)=b) =1$.  Expression \reff{n:2.49a} follows from
   \reff{n:2.49} since we are dealing with birth and death chains.
   Therefore, combining \reff{n:2.48} and \reff{n:2.49a} we have
   \begin{equation}
   \label{n:2.50}
     \tau^{(S)}  \le   K_2 \, \max(\E_0 T_{d/2}, \E_d T_{d/2}
     ):=f_S(d).
   \end{equation}

   In general, a coupling implies an inequality like (\ref{classic}). However,
   the inequality becomes an equality for a certain maximal (non-Markovian)
   coupling, described by Griffeath  \cite{dg:75}.
   Let $T_{\bx}^1$ be the coupling time for the maximal coupling
   between $s(\bX(t))$ and $s(\bY(t))$
    such that
   \[ \| P_{\bx}(S(\bX(t)) = \cdot) - \pi_S(\cdot ) \| = \P(T_{\bx}^{1} >
   t).\]
   Then
   \[ d_S(t)=\max_{\bx \in Q_d}\| P_{\bx}(S(\bX(t)) = \cdot ) - \pi_S(\cdot) \| = \max_{\bx \in Q_d}\P(T_{\bx}^{1} >
   t).\]
   By the definition of $\tau^{(S)}$ it is clear that $P(T_{\bx}^1>\tau^{(S)})
   \le {1\over {2e}}$.  Moreover, by the ``submultiplicativity'' property (see
   Aldous and Fill \cite{da:jf:99}, chapter 2)
   \begin{equation}
   %\label{submul}
   d(s+t)\le 2d(s)d(t),~~~ s,t \ge 0,
   \end{equation}
   we have that
   \begin{equation}
   \label{submul2}
   P(T_{\bx}^1>k\tau^{(S)})\le {1\over {2e^k}},~~~ k\ge 1.
   \end{equation}

   Thus
   \begin{eqnarray}
    \E T^{1}_{\bx} & = & \sum_{k=1}^\infty \E \bigl(T^{1}_{\bx} \one((k-1)\tau^{(S)} <T^{1}_{\bx} \le k
   \tau^{(S)})\bigr) \nonumber \\
                    & \le & \tau^{(S)} \sum_{k=1}^\infty k P((k-1)\tau^{(S)}
   <T^{1}_{\bx} \le k \tau^{(S)}) \nonumber \\
                    & \le & \tau^{(S)} \sum_{k=1}^\infty k P((k-1)\tau^{(S)}
   <T^{1}_{\bx} ) \nonumber \\
       & \le & \tau^{(S)} \left( 1+\sum_{k=2}^\infty k {1\over {2e^{k-1}}}
   \right). \nonumber \label{etx}
   \end{eqnarray}
   Since the series in the right hand side converges we have
   \[  \E T^{1}_{\bx} =O(f_S(d)).\]

   Once the bivariate $\bS$ process hits the diagonal
   \begin{equation}
     \label{n:3}
     \bD = \{(\bx,\by) \in Q_d \times Q_d; \sum_{i=1}^d x_i
     =\sum_{i=1}^d y_i\},
   \end{equation}
   we devise one obvious coupling that forces the bivariate $\bX$ process to stay
   in $\bD$ and such that the distance defined in \reff{n:2.2} between the
   marginal processes does not
   decrease. In words: we select one coordinate at random; if the marginal
   processes coincide in that coordinate, we allow them to evolve together;
   otherwise we select another coordinate in order to force two new
   coincidences.  Formally,
   for each ($\bX(t),\bY(t)) \in \bD$, let $I_1, I_2$ and $I_3$ be the
   partition of $\{0,1, \ldots,d\}$ such that
   \begin{eqnarray*}
   I_1 & = & \{i; X_i(t) = Y_i(t)\} \\
   I_2 & = & \{i; X_i(t) = 0, Y_i(t) = 1\} \\
   I_3 & = & \{i; X_i(t) = 1, Y_i(t) = 0\}
   \end{eqnarray*}

   Given $(\bX(t),\bY(t)) \in \bD $, choose $i$ u.a.r.  from $\{0,1,
   \ldots, d\}$.
   \begin{itemize}
   \item[(a)] If $i \in I_1$;
   \begin{enumerate}
   \item If $X_i(t) = 1$ then make $X_i(t+1) = Y_i(t+1) = 0$ with
     probability $p_{s(\bX(t)) - 1}$; otherwise $X_i(t+1) = Y_i(t+1) =
     1$.
   \item If $X_i(t) = 0$ then make $X_i(t+1) = Y_i(t+1) = 1$ with
     probability $p_{s(\bX(t))}$; otherwise $X_i(t+1) = Y_i(t+1) =
     0$.
   \end{enumerate}
   \item[(b)] If $i \in I_2$;
   \begin{enumerate}
   \item Select $j \in I_3$ u.a.r.;
   \item Make $X_i(t+1) = Y_j(t+1) = 1$ with probability $p_{s(\bX(t))}$;
     otherwise $X_i(t+1) = Y_j(t+1) = 0$.
   \end{enumerate}
   \item[(c)] If $i \in I_3$;
   \begin{enumerate}
   \item Select $j \in I_2$ u.a.r.;
   \item Make $X_i(t+1) = Y_j(t+1) = 0$ with probability $p_{s(\bX(t))-1}$;
     otherwise $X_i(t+1) = Y_j(t+1) = 1$.
   \end{enumerate}
   \end{itemize}

   Then, it is easy to check that $(\bX(t+1),\bY(t+1)) \in \bD$ and
   $d(\bX(t+1),\bY(t+1)) \le d(\bX(t),\bY(t))$. Moreover,
   noticing that $|I_2| = |I_3| = d(\bX(t),\bY(t))/2$, we have
   \begin{equation}
   \label{n:10.1}
   \P\bigl(d(\bX(t+1),\bY(t+1)) = d(\bX(t),\bY(t)) - 2 \given \bX(t),
   \bY(t)\bigr) =   \frac{ d(\bX(t),\bY(t))}{2d} (p_{s(\bX(t))} +
   p_{s(\bX(t))-1})
   \end{equation}
   \begin{equation}
   \label{n:10.2}
   \P\bigl(d(\bX(t+1),\bY(t+1)) = d(\bX(t),\bY(t)) \given \bX(t),
   \bY(t)\bigr)  =  1 \, - \,  \frac{ d(\bX(t),\bY(t))}{2d}
   (p_{s(\bX(t))} + p_{s(\bX(t))-1}).
   \end{equation}

   In this case, it is straightforward to compute
   \begin{eqnarray}
     m(i,s(\bX(t))) & = & i - \E \bigl[d(\bX(t+1),\bY(t+1)) \given
     d(\bX(t),\bY(t))=i, \bX(t), \bY(t)  \bigr]  \nonumber \\
          & = &  \frac{i}{d} (p_{s(\bX(t))} + p_{s(\bX(t))-1}) \label{n:19.1}.
   \end{eqnarray}

   Let $T_{\bx}^2$ be the coupling time for the second coupling just described.
   That is, let $T_{\bx}^2=\inf\{t>T_{\bx}^1: d(\bX(t),\bY(t))=0\}$. Then,
as a consequence of the optional sampling theorem for martingales we have
   the following comparison lemma ({\it cf.} Aldous and Fill
   \cite{da:jf:99}, Chapter 2).

   \begin{lemma}
   \begin{equation}
     \label{n:12}
     \E\bigl[T_{\bx}^{2}| d(\bX(T_{\bx}^{1}),\bY(T_{\bx}^{1})) = L,
     (\bX(T_{\bx}^{1}),\bY(T_{\bx}^{1})) \in \bD, s(\bX(T_{\bx}^{1}))=
     s\bigr] \, \le \,
     \sum_{i=1}^{L} \frac{d}{i(p_{s} + p_{s-1})}
   \end{equation}
   for all $s=0,1,\ldots,d$.
   \end{lemma}
   \proof Define $(\bX'(t), \bY'(t)) = (\bX(t+T_{\bx}^{1}),
   \bY(t+T_{\bx}^{1})$ for all $t \ge 0$. Define $Z(t) = d(\bX'(t),
   \bY'(t))$ and $\FF_t = \sigma(\bX'(t), \bY'(t))$. Then, it follows from
   \reff{n:19.1} that
   \begin{equation}
   \label{n:19.2}
   m(i,s) = i-E\bigl[Z(1)| Z(0) = i, s(\bX'(0))=s \bigr].
   \end{equation}
   Also, for all $s \in \{1, \ldots, d-1\}$, $0 < m(1,s) \le m(2,s) \le
   \ldots \le m(d,s)$.  Fix $s \in  \{1, \ldots, d-1\}$ and write
   \begin{equation}
   \label{n:19.3}
   h(i) = \sum_{j=1}^{i} \frac{1}{m(i,s)}
   \end{equation}
   and extend $h$ by linear interpolation for all real $0 \le x \le
   d$. Then $h$ is concave and for all $i \ge 1$
   \begin{eqnarray*}
   \E \bigl[ h(Z(1)) \given  Z(0)=i, s(\bX'(0))=s \bigr] & \le & h(i -
   m(i,s)) \\
     & \le & h(i) - m(i,s) h'(i) \\
     & = & h(i) - 1,
   \end{eqnarray*}
   where the first inequality follows from the concavity of $h$ and $h'$ is
   the first derivative of $h$. Now, defining $\tilde{h}$ such that
   \begin{equation}
   \label{n:19.4}
   h(i) = 1 + \sum_{j} \P \bigl[ h(Z(1)) \given  Z(0)=i, s(\bX'(0))=s
   \bigr] h(j) + \tilde{h}(i)
   \end{equation}
   and
   \begin{equation}
     \label{n:19.5}
     M(t) = t + h(Z(t)) + \sum_{u=0}^{t-1} \tilde{h}(Z(u))
   \end{equation}
   we have that $M$ is an $\FF_t$-martingale and applying the optional
   sampling theorem to the stopping time $T_0 = \inf\{t; Z(t)=0\}$ we have
   \begin{equation}
     \label{n:19.6}
     \E \bigl[ M(T_0)\given  Z(0)=i, s(\bX'(0))=s\bigr] =  \E \bigl[
     M(0)\given  Z(0)=i, s(\bX'(0))=s\bigr] = h(i).
   \end{equation}

   Noticing that $M(T_0) \ge T_0$ and $T_0 = T_{\bx}^{2}$, we obtain
 the desired result $\bullet$

   Since $s(\bX(t))$ is distributed as $\pi_S$, we can write:
   \begin{equation}
     \label{n:13}
     \E\bigl[T_{\bx}^{2}| d(\bX(T_{\bx}^{1}),\bY(T_{\bx}^{1})) = L,
     (\bX(T_{\bx}^{1}),\bY(T_{\bx}^{1})) \in \bD
     \bigr] \, \le \,
     \sum_{s=0}^d \pi_S(s)\sum_{i=1}^{L} \frac{d}{i(p_{s} + p_{s-1})}:=g(d).
   \end{equation}

   Putting the pieces together, we have found a coupling time $T_{\bx}$ for the
   whole process such that
   $$\E T_{\bx} \le f_S(d)+g(d).$$
   The task now is to find explicit bounds for $f_S(d)$ and $g(d)$ for particular
   workable cases.

   To avoid unnecessary complications, we will
   assume $d=2m$, and compute only the hitting times for the $\bS$ process of the
   type $E_0 T_m$.  Hitting times in birth-and-death processes assume the
   following
   closed-form  (see \cite{jp:pt:96} for an electrical derivation):

   $$\E_kT_{k+1}={1\over {\pi_kP(k,k+1)}}\sum_{i=0}^k\pi_i,~~~~~0\le k\le
   d-1,$$
   and in our case this expression turns into
   $$\E_kT_{k+1}={{\sum_{i=0}^k{d\choose i}}\over {{{d-1}\choose
         k}p_k}}.$$
   Therefore
   \begin{equation}
   \label{medio}
   \E_0T_m=\sum_{k=0}^{m-1}{{\sum_{i=0}^k{{2m}\choose i}}\over {{{2m-1}\choose
         k}p_k}}.
   \end{equation}

   (i) In case all $p_k=1$, we have the simple random walk on the cube,
   and it turns out there is an even more compact expression of
   (\ref{medio}), namely:
   \begin{equation}
   \label{blom}
   \E_0T_m=\sum_{k=0}^{m-1}{{\sum_{i=0}^k{{2m}\choose i}}\over {{{2m-1}\choose
         k}}}=m\sum_{k=0}^{m-1}{1\over {2k+1}},
   \end{equation}
   as was proved in  \cite{gb:89}, and the right hand side of (\ref{blom}) equals
   $\displaystyle m[H(2m)-{1\over 2}H(m)]$, where $H(n)=1+{1\over
     2}+\cdots +{1\over n}$,
    allowing us to conclude
   immediately that in this case $\E_0T_m=\E_0T_{d/2}\approx {d\over 4}\log d+{d
     \over 4}\log 2$.
   \vskip .1 in
   Also, we have
   \begin{equation}
     \label{n:12.1}
     \E[T_{\bx}^{2}| d(\bX(T_{\bx}^{1}),\bY(T_{\bx}^{1})) = L,
     (\bX(T_{\bx}^{1}),\bY(T_{\bx}^{1})) \in \bD] \le \frac{d}{2p} \sum_{i=1}^{L}
     \frac{1}{i} \approx  \frac{d}{2p} O(\log L).
   \end{equation}
   Thus in this case both $f_S(d)$ and $g(d)$, and a fortiori $\E T_{\bx}$ and
   $\tau$, are $O(d \log d)$.

   (ii) For the Aldous cube, $p_k={k\over {d-1}}$, and (\ref{medio})
   becomes (recall this cube excludes the origin):
   \begin{equation}
   \label{ald}
   \E_1T_m=\sum_{k=1}^{m-1}{{\sum_{i=0}^k{{2m}\choose i}}\over {{{2m-2}\choose
         {k-1}}}}=\sum_{k=0}^{m-2}{{\sum_{i=0}^k{{2m}\choose i}}\over
         {{{2m-2}\choose k}}}+\sum_{k=0}^{m-2}{{{2m}\choose {k+1}}\over
         {{{2m-2}\choose k}}}.
   \end{equation}

   After some algebra, it can be shown that the second summand in (\ref{ald})
   equals
   $\displaystyle (2m-1)[H(2m-1)-{1\over m}]$,
   and the first summand can be bounded by
   twice the expression in (\ref{blom}), on account of the fact that
   $\displaystyle {{2m-1}\choose k}\le 2 {{2m-2}\choose k},$
   for $0\le k\le m-1$.  Therefore, we can write

   \[\E_1T_{d/2}\le {3\over 2} d\log d+\mbox{ smaller~ terms},\]
   thus improving by a factor of ${1\over 2}$ the computation of the same hitting
   time in  \cite{fc:rg:97}.

   Also, we have
   \begin{equation}
     \label{n:12.2}
     \E[T_{\bx}^{2}| d(\bX(T_{\bx}^{1}),\bY(T_{\bx}^{1})) = L,
     (\bX(T_{\bx}^{1}),\bY(T_{\bx}^{1})) \in \bD, s(\bX(T_{\bx}^{1}))  =s] \le
     \sum_{i=1}^{L} \frac{d(d-1)}{i(2s - 1)}
   \end{equation}

   Thus, in this case
   \begin{eqnarray}
    \lefteqn{ \E[T_{\bx}^{2}| d(\bX(T_{\bx}^{1}),\bY(T_{\bx}^{1})) = d,
     (\bX(T_{\bx}^{1}),\bY(T_{\bx}^{1})) \in \bD]}\nonumber  \\
    & \quad \quad  \le & \sum_{s=1}^{d} \pi_{S}(s) \sum_{i=1}^{d}
     \frac{d(d-1)}{i(2s - 1)} \nonumber  \\
    & \quad \quad  \le & \Phi(-\sqrt{d}/3)\sum_{i=1}^{d} \frac{d(d-1)}{i} +
    ( 1 - \Phi(-\sqrt{d}/3))  \sum_{i=1}^{d} \frac{d(d-1)}{i(2d/3 -
     1)} \nonumber \\
     & \quad \quad  \approx & e^{-d/9} d(d-1) \log d + (1- e^{-d/9}) d \log
     d.  \label{n:23.2}
   \end{eqnarray}

   And so $\tau=O(d \log d)$ also in  this case.

   (iii) Slower walks.  Consider the case when the probability $p_k$
   grows exponentially in $k$, more specifically
   \begin{equation}
     \label{n:19.7}
     p_k = \Bigl( \frac{k+1}{n+1} \Bigr)^{\alpha}
   \end{equation}
   with $\alpha > 1$. In this case, it seems that \reff{medio} is useless
to get a
   closed expression for $\E_0T_{d/2}$. However, Graham and Chung  
\cite{fc:rg:97}
   provide the following bound
   \begin{equation}
   \label{n:19.8}
     \E_i T_{d/2} \le c_0(\alpha) d^{\alpha}, \quad \mbox{ for all } d
     \ge d_0(\alpha), 0 \le i \le d
   \end{equation}
   where $ c_0(\alpha)$ and $d_0(\alpha)$ are constants depending only on
   $\alpha$. Moreover, \reff{n:13} becomes
   \begin{eqnarray}
   g(d) & = & \sum_{s=0}^{d} \pi_{S}(s) \sum_{i=1}^{d}
   \frac{d(d+1)^{\alpha}}{i((s+1)^{\alpha} + s^{\alpha})} \nonumber \\
        & = & d(d+1)^{\alpha} \sum_{s=0}^{d}
        \frac{\pi_{S}(s)}{(s+1)^{\alpha} + s^{\alpha} } \sum_{i=1}^{d}
        \frac{1}{i} \nonumber \\
        & \approx &  d(d+1)^{\alpha} \log d \sum_{s=0}^{d}
        \frac{\pi_{S}(s)}{(s+1)^{\alpha} + s^{\alpha}}\nonumber\\
& \le &d(d+1)^{\alpha} \log d  \sum_{s=0}^{d}
        \frac{\pi_{S}(s)}{(s+1)^{\alpha}} \nonumber\\
& = & d(d+1)^{\alpha} \log d  \E \Bigl[ \frac{1}{(1+X)^{\alpha}}\Bigr],
 \label{n:22.1}
   \end{eqnarray}
where $X$ is a Binomial ($d, {1\over 2}$) random variable.  Jensen's
inequality and the same argument that lead to \reff{n:23.2} show that
$\E[(1+X)^{-\alpha}] \sim O(d^{-\alpha})$ and  \reff{n:22.1} can be
bound by $O(d \log d)$. This fact  together with \reff{n:19.8} allows
us to conclude that  $\tau=O(d^{\alpha })$ in this case, thus
improving on the rate of the mixing  time provided in \cite{fc:rg:97}
by a factor of $\log d$. 
\vskip .2 in

   \noindent {\bf Acknowledgments. } This paper was initiated when both
   authors were visiting scholars at the Statistics Department of the
   University of California at Berkeley. The authors wish to thank their 
hosts, especially Dr. David Aldous with whom they shared many fruitful 
discussions.  This work was partially  supported by FAPESP 99/00260-3.

   \end{document}